\documentclass[a4paper]{jpconf}
\usepackage{graphicx}
\usepackage{amsmath}

\begin{document}
\title{Three-body losses of repulsively interacting three-component fermionic atoms in optical lattices}

\author{Sei-ichiro Suga$^{1}$ and Kensuke Inaba$^{2,3}$}

\address{$^1$Graduate School of Engineering, University of Hyogo, Himeji 671-2280, Japan}
\address{$^2$ NTT Basic Research Laboratories, NTT Corporation, Atsugi 243-0198, Japan}
\address{$^3$ JST, CREST, Sanbancho, Chiyoda-ku, Tokyo 102-0075, Japan}

\ead{suga@eng.u-hyogo.ac.jp}

\begin{abstract}
We investigate the effects of a repulsive three-body interaction on the Mott transition of the repulsively interacting three-component fermionic atoms in optical lattices by means of the self-energy functional approach.
We find that the three-body repulsion hardly affects the qualitative features of the Mott transition, because the three-body repulsion does not compete with the two-body repulsions.
When the three-body repulsion is extremely strong, the triple occupancy vanishes in the Fermi liquid state.
This situation is equivalent to that caused by strong three-body losses.
Our results imply that three-body losses have little influence on the Mott transitions in the repulsively interacting three-component fermionic atoms in optical lattices.
\end{abstract}

\section{Introduction}
High controllability of cold atoms in optical lattices paves a way for studying novel quantum many-body systems that are difficult to realize in condensed matter systems.
Actually, degenerate fermionic atoms with three-component internal degrees of freedom were actively investigated and their interesting features have been revealed in recent experiments \cite{Ottenstein2008,Huckans2009}.
For three-component (color) fermionic atoms in optical lattices, theoretical studies based on unperturbative methods have pointed out various interesting features.
For attractively interacting systems, a phase transition appears between the color superfluid and the trionic state, namely, the singlet bound state of three-color atoms  \cite{Rapp2007,Inaba2009a,Inaba2011}.
For repulsively interacting systems, a characteristic Mott phase called the paired Mott insulator (PMI) appears at non-integer half filling \cite{Inaba2010b,Suga2011,Inaba2013}, and also a superfluid phase appears in the vicinity of half filling \cite{Inaba2012,Inaba2013}.
However, most of these previous studies have not dealt with three-body losses, which intrinsically exist in the realistic cold atom systems.
Some recent theoretical studies considering thee-body losses for the attractively interacting system have suggested that strong three-body losses induce a phase separation instead of a trion formation \cite{Priv,Titv}.

In this paper, we discuss how three-body losses affect the PMI transition in repulsively interacting three-component fermionic atoms in optical lattices.
For this purpose, we investigate the three-component fermionic Hubbard-type model including a repulsive three-body interaction term.
We find that the extremely strong three-body repulsion projects out the triple occupancy effectively. This situation is equivalent to that caused by very large three-body losses \cite{Priv,Titv}.

\section{Model and method}
According to the conventional model for cold atoms in optical lattices \cite{Jaksch}, we include only the nearest-neighbor tunneling and set the on-site interaction between different colors. 
The model Hamiltonian reads as
\begin{eqnarray}
{\cal H}=-t \sum_{\langle i,j \rangle}\sum_{\alpha=1}^{3}
       \hat{a}^\dag_{i\alpha} \hat{a}_{j\alpha}
  - \sum_{i}\sum_{\alpha=1}^{3} \mu_\alpha \hat{n}_{i \alpha}
  + \frac{1}{2}\sum_{i}\sum_{\alpha\not=\beta}
       U_{\alpha\beta} \hat{n}_{i \alpha} \hat{n}_{i \beta}
  + V\sum_{i} \hat{n}_{i1} \hat{n}_{i2} \hat{n}_{i3},
\label{eq_model}
\end{eqnarray}
where the subscript $\langle i,j \rangle$ is the summation over the nearest-neighbor sites, and $\hat{a}^\dag_{i\alpha} (\hat{a}_{i\alpha})$ and $\hat{n}_{i\alpha}$ are creation (annihilation) and number operators of a fermion with color $\alpha(=1,2,3)$ at the $i$th site, respectively.
The first three terms describe the three-component Hubbard Hamiltonian as usual. The last term denotes the three-body interaction with $V>0$.
The chemical potentials $\mu_\alpha$ are adjusted to choose half filling with a balanced population of each color atoms ($\langle \hat{n}_{i \alpha}\rangle =1/2$),
leading to $\mu_\alpha=\left(U_{\alpha\beta}+U_{\gamma\alpha} \right)/2$. 
Thus, $3/2$ atoms exist at each site on the average at half filling.
For simplicity, we set $U_{12}\equiv U (>0)$ and $U_{23}=U_{31}\equiv U' (>0)$, and introduce parameters $v\equiv V/U'$ and $R\equiv U/U'$.
We focus on the Mott transition by neglecting the symmetry breaking phases and the effects of the confinement potential.

To take account of the local correlation effects precisely, we employ an unperturbative method, a self-energy functional approach (SFA) \cite{Potthoff03a,Potthoff03b}.
This method has been successfully applied to investigate Mott transitions at zero and finite temperatures in strongly correlated electron systems \cite{Potthoff03a,Potthoff03b,Inaba05}.
Furthermore, the SFA allows us to deal with superfluid transitions in the repulsively \cite{Inaba2012} and attractively \cite{Inaba2009a} interacting three-component fermionic atoms in optical lattices.
The SFA is based on the Luttinger-Ward variational principle $\partial\Omega/\partial \Sigma=0$, where $\Omega$ is the grand potential of the system and $\Sigma$ is the {\it physical} self-energy. 
We introduce a reference system ${\cal H}_{\rm ref}$ that has the same interaction terms as those of the original Hamiltonian (\ref{eq_model}), so that the grand potential $\Omega$ can be written as follows: 
\begin{eqnarray}
\Omega =\Omega_{\rm ref} +{\rm Tr}\ln
    \left[-(\omega+\mu-{\bf t}-{\boldsymbol \Sigma}_{\rm ref})^{-1}\right]
    -{\rm Tr}\ln \left[-(\omega+\mu-{\bf t}_{\rm ref}-{\boldsymbol \Sigma}_{\rm ref})^{-1}\right],
\label{eq:omega_SFA}
\end{eqnarray}
where $\Omega_{\rm ref}$ and ${\boldsymbol \Sigma}_{\rm ref}$ are the grand potential and the self-energy of the reference system ${\cal H}_{\rm ref}$, and 
${\bf t}_{\rm ref}$ and ${\bf t}$ are parameter matrices of the one-body terms of the reference Hamilitonial and the original Hamiltonian, respectively.
By choosing the parameter matrix ${\bf t}_{\rm ref}$ to satisfy the variational condition
$\partial\Omega/\partial {\bf t}_{\rm ref}= \left(\partial\Omega/\partial {\boldsymbol \Sigma}_{\rm ref} \right) \left( \partial {\boldsymbol \Sigma}_{\rm ref}/\partial {\bf t}_{\rm ref} \right)=0$,
we obtain a reference self-energy ${\boldsymbol \Sigma}_{\rm ref}$ that properly reproduces the physical properties of the original system.

We use the following local reference Hamiltonian: ${\cal H}_{\rm ref}=\sum_i{\cal H}_{\rm ref}^{(i)}$, 
\begin{eqnarray}
{\cal H}_{\rm ref}^{(i)}&=&\sum_{\alpha=1}^3 \left( \epsilon_{c\alpha}\hat{c}^\dag_{i\alpha} \hat{c}_{i\alpha}+ \epsilon_{a\alpha}\hat{a}^{\dag}_{\alpha}\hat{a}_{\alpha}\right)
   +\sum_{\alpha=1}^3\left(\Gamma_{\alpha} \hat{c}^\dag_{i\alpha}\hat{a}_{\alpha}+H.c.\right) \nonumber\\
&& \ \ \ \ \ \ \ \ \ \ \ \ \ \ \ \ \ \ \ \ \ \ \ \ \ \ \ \ \ \
+\frac{1}{2}\sum_{\alpha\not=\beta=1}^3 U_{\alpha\beta} \hat{n}_{i \alpha} \hat{n}_{i \beta}  + V\sum_{i} \hat{n}_{i1} \hat{n}_{i2} \hat{n}_{i3},
\label{eq_ref_model}
\end{eqnarray}
where $\hat{a}^{\dag}_{\alpha}(\hat{a}_{\alpha})$ is the creation (annihilation) operator of a noninteracting fermion with color $\alpha$ connecting to the $i$th site in the original lattice. 
Here, the variational parameters are $\epsilon_{c\alpha}$, $\epsilon_{a\alpha}$, and $\Gamma_{\alpha} \; (\alpha=1,2,3)$. 
In the present local approximation, the hybridization ($\Gamma_{\alpha}$) between the noninteracting fermions and the original lattice fermions effectively captures the itineracy of each color atoms. 
By searching for the optimized parameters $\epsilon_{c\alpha}$, $\epsilon_{a\alpha}$, and $\Gamma_{\alpha}$, we can thus discuss characteristic Mott transitions in the three-component repulsive fermionic atoms with the three-body repulsion in optical lattices.



In the following, for noninteracting atoms, we use a semi-circular density of states, $\rho_0(x)=\sqrt{4t^2-x^2}/(2\pi t^2)$, which is independent of color. 
We calculate the quasiparticle weight $Z_{\alpha}=[1-\partial \Sigma_{{\rm ref}, \alpha}(\omega)/\partial \omega]|_{\omega=0}^{-1}$, using the reference self-energy $\Sigma_{{\rm ref}}$. 
We also calculate 
the double occupation $D_{\alpha\beta}= \partial\Omega/ \partial U_{\alpha\beta} = \langle \hat{n}_{i \alpha} \hat{n}_{i \beta} \rangle$ and the triple occupation
$T_{\rm occ}= \partial\Omega/ \partial V = \langle \hat{n}_{i1} \hat{n}_{i2} \hat{n}_{i3} \rangle$, where
$D_{\alpha\beta}$ and $T_{\rm occ}$ are independent of the site index $i$.
Note that $D_{\alpha\beta}=1/4$ and $T_{\rm occ}=1/8$ for noninteracting atoms.
Because of $U_{12}\not= U_{23}=U_{31}$, $Z_1=Z_2$ and $D_{13}=D_{23}$ are satisfied.
The hopping integral $t$ is used in units of energy.

\section{Results}
\begin{figure}
\begin{center}
\includegraphics[width=14.0cm]{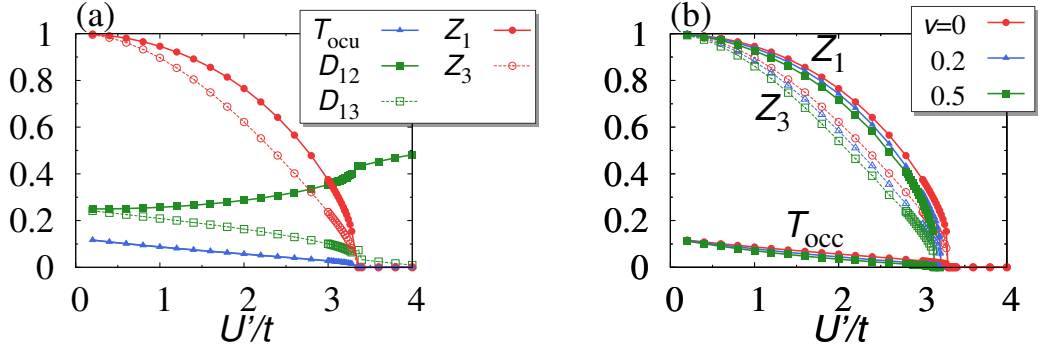}
\end{center}
\caption{(Color online) (a) Quasiparticle weight $Z_{\alpha}$, double occupation $D_{\alpha\beta}$, and triple occupation $T_{\rm occ}$ for $v=0$ at half filling. (b) Quasiparticle weight $Z_{\alpha}$  and triple occupation $T_{\rm occ}$ for $v=0, 0.2$, and $0.5$ at half filling, where filled and open symbols for $Z_{\alpha}$ represent $Z_1(=Z_2)$ and $Z_3$, respectively.
These calculations are performed for $R=0.1$ at zero temperature. Lines are guide to eyes.
}
\label{fig1}
\end{figure}
We first analyze the system without the three-body interaction ($V=0$) for $R=0.1$ at zero temperature.
Figure \ref{fig1}(a) exhibits that $Z_1$, $Z_3$, and $D_{13}$ become zero at $U'_c/t\sim3.4$, while $D_{12}$ approaches the maximum $D_{12}=1/2$ in $U'>U'_c$. This is a manifestation of the quantum phase transition to the PMI, where pairs of color-1 and 2 atoms are formed to avoid the two stronger repulsions.
This Mott transition at non-integer half filling occurs as a result of the pair formation.
Here, the total number of the effective particles at each site becomes an integer as $n_{\rm pair}+n_3=1$, where $n_{\rm pair}$ is the average number of the paired particles, and $n_3$ is that of the color-3 atoms \cite{Inaba2010b}.
We also find that the triple occupation $T_{\rm occ}$ decreases with increasing $U'/t$ and becomes zero at the PMI transition point.

Next, we discuss the effects of a weak three-body interaction for $v=0.2$ and $0.5$.
Figure \ref{fig1}(b) shows that, as $v$ is increased, $Z_1$ and $Z_3$ decrease and the critical value of the PMI transition shifts to smaller $U'/t$.
The triple occupation $T_{\rm occ}$ also reduces with increasing $v$.
Importantly, no qualitative difference is observed between the $v=0$ and $v\not=0$ cases.
This is because the repulsive three-body and two-body interactions do not compete with each other.

The decrease in the triple occupation allows us to deal with the repulsive three-body interaction by using the mean-field approximation.
The mean-field three-body interaction reduces to $ \sim  
V\left(\langle \hat{n}_{i1}\rangle \hat{n}_{i2} \hat{n}_{i3} + \langle \hat{n}_{i2}\rangle \hat{n}_{i1} \hat{n}_{i3} + \langle \hat{n}_{i3}\rangle \hat{n}_{i1} \hat{n}_{i2}\right)$.
This suggests that the three-body repulsion can be renormalized into the two-body repulsions $U$ and $U'$.
Using the condition $\langle \hat{n}_{i\alpha} \rangle =1/2$,
the renormalized two-body interactions are evaluated as $\tilde{U}=(R+v/2)U'$ and $\tilde{U'}=(1+v/2)U'$.
Thus, the slight shift of the PMI transition point with increasing $v$ shown in Fig. \ref{fig1}(b) can be qualitatively explained by this mean-field approximation.
\begin{figure}[t]
\begin{center}
\includegraphics[width=14.0cm]{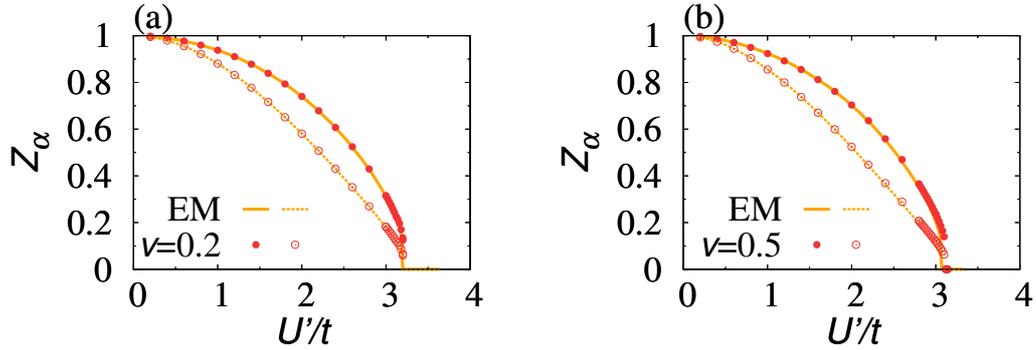}
\end{center}
\caption{ (Color online)  Quasiparticle weight $Z_\alpha$ derived from the effective model (EM) in Eq. (2) and from the original model in Eq. (1) for (a) $v=0.2$ and (b) $v=0.5$ at half filling. Filled and open symbols represent $Z_1(=Z_2)$ and $Z_3$, respectively, obtained from the original model, while solid and broken lines represent those obtained from the effective model. Calculations are performed for $R=0.1$ at zero temperature.
}
\label{fig2}
\end{figure}
To discuss the validity of this mean-field approximation, we further perform numerical calculations on the basis of the effective three-component Hubbard model with the renormalized parameters $\tilde{U}$ and $\tilde{U'}$. The effective model reads as
\begin{eqnarray}
{\cal H}_{\rm eff}=-t \sum_{\langle i,j \rangle}\sum_{\alpha=1}^{3}
       \hat{a}^\dag_{i\alpha} \hat{a}_{j\alpha}
  - \sum_{i}\sum_{\alpha=1}^{3} \mu_\alpha \hat{n}_{i \alpha}
  + \sum_{i}\left( \tilde{U}\hat{n}_{i1}\hat{n}_{i2}+\tilde{U'}\hat{n}_{i2}\hat{n}_{i3}+\tilde{U'}\hat{n}_{i1}\hat{n}_{i3} \right).
\end{eqnarray}
We compare $Z_1$ and $Z_3$ obtained from this effective model with those obtained from the original model.
Figure \ref{fig2} shows that the effective model well reproduces the feature of the original one quantitatively for both $v=0.2$ and $v=0.5$, indicating that the mean-field approximation well captures the effects of the weak three-body repulsion.

\begin{figure}[t]
\begin{center}
\includegraphics[width=14.0cm]{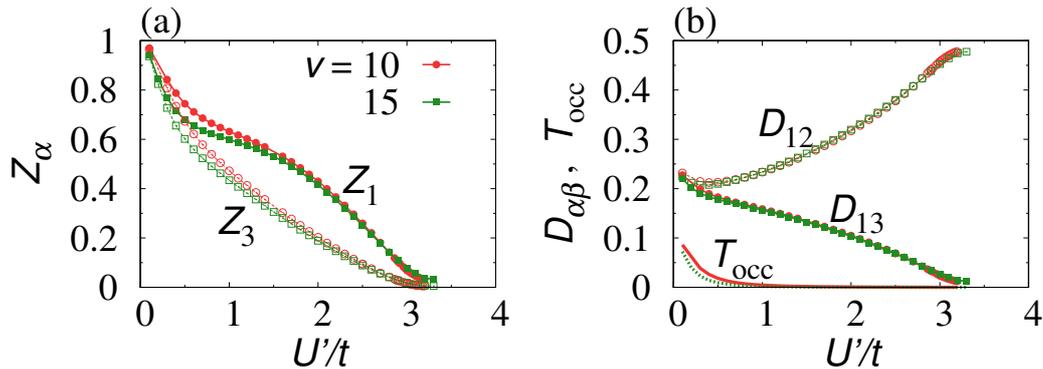}
\end{center}
\caption{ (Color online) (a) Quasiparticle weight $Z_\alpha$, and (b) double occupation $D_{\alpha\beta}$ and triple occupation $T_{\rm occ}$ for $v=10$ and $v=15$ at half filling.
Calculations are performed for $R=0.1$ at zero temperature. Lines are guide to eyes.
}
\label{fig3}
\end{figure}
We further investigate the effects of an extremely strong three-body repulsion.
We show the quasiparticle weights $Z_1$ and $Z_3$ in Fig. \ref{fig3}(a),  and the double and triple occupations in Fig. \ref{fig3}(b) for $v=10$ and $v=15$.
We find that the triple occupation decreases quickly and vanishes at $U'/t \sim 1.5$, where the system is in the Fermi liquid. This behavior of $T_{\rm occ}$ is in contrast with that in the weak $v$ shown in Fig. \ref{fig1}(b), where the $T_{\rm occ}$ vanishes at the PMT transition point. For $U'/t > 1.5$, the quasiparticle weights and double occupations are almost independent of $v$.
These findings for extremely strong three-body repulsions mean that the properties close to the PMI transition point can be described by the constrained Hamiltonian $\hat{\cal P}^\dag{\cal H}\hat{\cal P}$, where $\hat{\cal P}$ is a projection operator to exclude the triple occupancy at any site \cite{Priv,Titv}.
Significantly, the PMI transition still occurs under the extremely strong three-body repulsion, which can be confirmed from the behavior that
$Z_1$ and $Z_2$ go to zero, and $D_{12}$ and $D_{13}$ approach $1/2$ and $0$, respectively, towards the transition point $U'_{\rm c}/t \sim 3.2$.
This result also originates form the absence of the competition between the repulsive three-body and two-body repulsions.

\section{Discussion and conclusion}
We have shown that the repulsive three-body interaction hardly influences the qualitative features of the PMI transition, because the three-body repulsion does not compete with the two-body repulsions.
For the extremely strong three-body repulsion, we find that the triple occupancy is projected out close to the PMI transition point.
This situation is equivalent to the system with very large three-body losses \cite{Priv,Titv}.
Therefore, we conclude that large three-body losses have little influence on the PMI transition in the repulsively interacting three-component fermionic atoms in optical lattices.
This conclusion is in contrast with that obtained in the attractively interacting three-component fermionic atoms, where the trionic state becomes unstable due to three-body losses, and the phase separation occurs towards paired and unpaired atoms \cite{Priv,Titv}. Its driving force is the competition between the repulsive three-body interaction and the attractive two-body interactions.

We finally comment on the quantum degenerate three-component fermionic ${\rm ^6Li}$ gases, which have been intensively investigated \cite{Ottenstein2008}.
In this system, the small three-body loss rate was observed in the repulsively interacting region under 550G-600G magnetic field, while the very large loss rate was observed in the attractive region around 500G \cite{Ottenstein2008}.
In the repulsive region,  in addition to the small loss coefficient, the two-body repulsive interactions itself decrease the triple occupation $T_{\rm occ}$ as shown in Fig. \ref{fig1}(a), which implies the suppression of three-body losses.
Note that the decrease in the number of color-$\alpha$ atoms per site caused by three-body losses is described by $dn_\alpha/dt \propto -\gamma T_{\rm occ}$, where $\gamma$ is a (positive) loss coefficient that includes the effects of the bare three-body loss coefficient $K_3$ \cite{Ottenstein2008} and the overlap integration $\int d{\bf r} |W({\bf r})|^6$ of the Wannier function $\left(  W({\bf r}) \right)$ of atoms in the lattice.
We have shown that $T_{\rm occ}$ is negligibly small close to the PMI transition point. We thus consider that the $^6$Li atom system in optical lattices is a possible candidate for observing the PMI.

\section*{Acknowledgments}
This work was supported by Grants-in-Aid for Scientific Research (C) (No. 23540467) and (B) (No. 25287104) from the Japan Society for the Promotion of Science.

\end{document}